\pdfoutput=1

\documentclass[twocolumn,english,superscriptaddress]{revtex4}
\usepackage{times,amssymb,amsmath,graphicx}
\usepackage[bookmarks=false,pdffitwindow=false,pdfstartview={FitH}]{hyperref}
\usepackage[T1]{fontenc}
\usepackage{babel}
\usepackage{color}

\begin{document}


\title{
  Thermoelectric properties of the Corbino disk in graphene
}

\author{Adam Rycerz\footnote{Corresponding author; e-mail:
  \href{mailto:rycerz@th.if.uj.edu.pl}{rycerz@th.if.uj.edu.pl}.}}
\affiliation{Institute for Theoretical Physics,
  Jagiellonian University, \L{}ojasiewicza 11, PL--30348 Krak\'{o}w, Poland}

\author{Katarzyna Rycerz}
\affiliation{Institute of Computer Science, AGH University of Science
  and Technology, al.\ Mickiewicza 30, PL--30059 Krak\'{o}w, Poland}

\author{Piotr Witkowski}
\affiliation{Institute for Theoretical Physics,
  Jagiellonian University, \L{}ojasiewicza 11, PL--30348 Krak\'{o}w, Poland}

\date{June 2, 2023}

\begin{abstract}
Thermopower and the Lorentz number for an edge-free (Corbino) graphene
disk in the quantum Hall regime is calculated within the Landauer-B\"{u}ttiker
formalism. By varying the electrochemical potential, we find that
amplitude of the Seebeck coefficient follows a~modified Goldsmid-Sharp
relation, with the energy gap defined by the interval between the zero
and the first Landau levels in bulk graphene. An analogous relation for
the Lorentz number is also determined.
Thus, these thermoelectric properties are solely defined by the magnetic
field, the temperature, the Fermi velocity in graphene,
and fundamental constants including the electron charge, 
the Planck and Boltzmann constants, being independent on geometric
dimensions of the system.
This suggests that the Corbino disk in graphene may operate
as a~thermoelectric thermometer, allowing to measure small temperature
differences between two reservoirs, if the mean temperature
and the magnetic field are known.  
\end{abstract}

\maketitle


\section{Introduction}
Almost twenty years after the discovery of graphene \cite{Nov04,Nov05,Zha05},
a~two-dimensional form of carbon hosting ultrarelativistic effective
quasiparticles \cite{Nov05,Zha05}, which forced researchers to re-examine
numerous effects previously known from mesoscopic physics
\cite{Das11,Roz11,Kat20,Lee20,TLi20,Ron21,Sch23}, 
it seems that the development of quantum Hall resistance standards
\cite{Kal14,Laf15,Kru18} can be considered as the most important 
application to emerge from this field of research.
Although the Hall bar setup is most commonly used (also in the study
of artificial graphene analogues \cite{Pol13,Mat16,Tra21}), 
the edge-free Corbino
geometry is often considered when discussing fundamental aspects of graphene
\cite{Kat20,Che06,Ryc09,Ryc10,Pet14,Abd17,Zen19,Sus20,Kam21,Yer21}.   
In such a~geometry, magnetotransport at high fields is unaffected by edge
states, allowing one to probe the bulk transport properties 
\cite{Zen19,Sus20,Kam21}. 
Recently, sharp resonances in the longitudinal conductivity, associated with
Landau levels, have been observed in graphene disks on hexagonal
boron nitride \cite{Zen19}. 

In addition to conductivity measurements, thermoelectric phenomena including
Seebeck and Nerst effects in graphene
\cite{Dol15,Wan11,Chi16,Mah17,Sus18,Sus19,Zon20,Dai20,Jay21,Cie22} and
other two-dimensional systems \cite{Lee16,Sev17,Qin18,DLi20} have
been studied thoroughly, providing valuable insights into the details
of the electronic structure of these materials.
In particular, for systems with a~wide bandgap $E_g$, the maximum
absolute value of the Seebeck coefficient can be approximated by
a~Goldsmid-Sharp value \cite{Hao10,Gol99}
\begin{equation}
\label{smaxgs}
  |S|_{\rm max}\approx{}\frac{E_g}{2eT},\ \ \ \ \ 
  \text{for }\ \ E_g\gg{}k_BT,  
\end{equation}
with the absolute temperature $T$, the electron charge $-e$, and
the Boltzmann constant $k_B$. (For more accurate approximations, see
Ref.\ \cite{Sus19}.)

The thermoelectric properties of graphene disks at zero (or low) magnetic
fields have also been considered \cite{Ryc21a,SLi22}. 
In the quantum Hall regime, thermoelectricity has been studied for
GaAs/AlGaAs based Corbino disks which host a~two-dimensional gas of
non-relativistic electrons \cite{Bar12,Kob13,dAm13,Rea20}.
However, analogous studies for graphene disks are missing so far. 

\begin{figure*}[!t]
  \includegraphics[width=\linewidth]{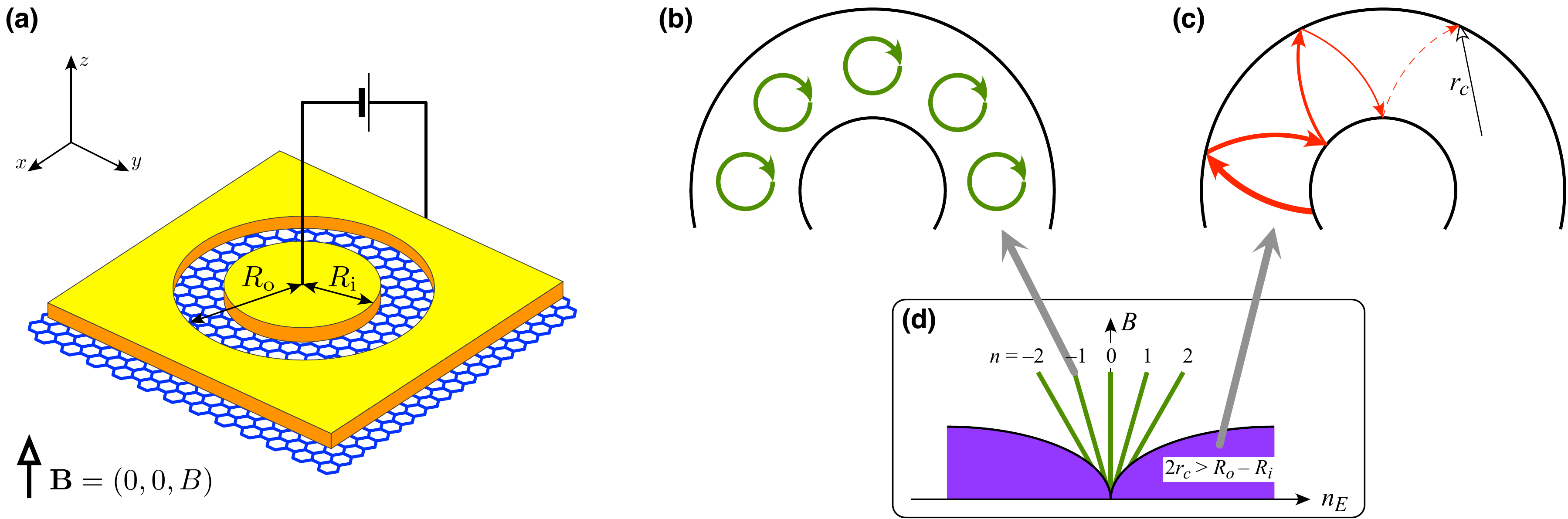}
  \caption{ \label{setup3:fig}
    (a) The Corbino setup in graphene. Voltage source passes the current
    between the circular leads (yellow areas) via disk-shaped sample with
    the inner radii $R_{\rm i}$ and the outer radii $R_{\rm o}$.
    Uniform magnetic field ${\bf B}=(0,0,B)$ is perpendicular to the sample. 
    Additional gate electrode (not shown) tunes the doping in the disk area.
    (b)--(d) Transport regimes for different fields and carrier
    concentrations $n_E$. At high fields, if doping is adjusted to a~Landau
    level ($E=E_{n{\rm LL}}$, $n=0,\pm{}1,\dots$) resonance occurs (b).
    At low field but high doping (such that the cyclotron diameter
    $2r_c>R_{\rm o}-R_{\rm i}$), incoherent scattering along classical
    trajectory governs the transport (c). 
  }
\end{figure*}

In this paper, we present numerical results on the Seebeck
coefficient and the Lorentz number (quantifying the ratio of thermal to
electrical conductivity) for the ballistic disk in graphene
(see Fig.\ \ref{setup3:fig}).
The results show that although the deviations from Eq.\ (\ref{smaxgs}) are
noticeable, the thermopower amplitude (determined by varying the doping
at fixed temperature $T$ and field $B$) can still be truncated by
a~closed-form function of the quantity $\Delta{}E_{\rm max}/(2eT)$, where
$\Delta{}E_{\rm max}\propto{\sqrt{B}}$ is the maximum interval between
Landau levels (LLs), playing a~role of the transport gap. 
A~similar conclusion applies to the maximum Lorentz number. 
Unusual sequence of LLs in graphene leads to relatively high thermoelectric
response is expected for micrometer-size disks at moderate fields $B<0.5\,$T
and few Kelvin temperatures.
The effect of smooth potential profiles is also discussed, 
introducing the electron-hole asymmetry of the transport properties 
\cite{Ryc21b,Ryc22}.

The remaining parts of the paper are organized as follows.
In Sec.\ \ref{modmet} we present details of our numerical approach, which
can be applied either in the idealized case where the electrostatic potential
energy is a~piecewise-constant function of the distance from the disk center,
or in the more general case of smooth potentials.
Our numerical results for both cases are presented
in Sec.\ \ref{resdis}. The conclusions are given in Sec.\ \ref{conclu}.

\section{Model and methods}
\label{modmet}

\subsection{Scattering of Dirac fermions}
Our analysis starts with the wave equation for massless Dirac fermions
in graphene at energy $E$ and uniform magnetic field $B$, which
can be written as (for the $K$ valley)
\begin{equation}
  \label{direqvr}
  \left[
    v_F\,(\mbox{\boldmath$p$}+e\mbox{\boldmath$A$})\cdot
    \mbox{\boldmath$\sigma$} + V(r)
  \right]
  \Psi=E\Psi, 
\end{equation}
where $v_F=\sqrt{3}\,t_0a/(2\hbar)\approx{}10^6\,$m$/$s is the
energy-independent Fermi velocity in graphene (with $t_0=2.7\,$eV the
nearest-neighbor hopping integral and $a=0.246\,$nm the lattice parameter),
$\mbox{\boldmath$p$}=(p_x,p_y)$ is the in-plane momentum operator with
$p_j=-i\hbar{}\partial_j$, we choose the symmetric gauge
$\mbox{\boldmath$A$}=\frac{B}{2}(-y,x)$, and
$\mbox{\boldmath$\sigma$}=(\sigma_x,\sigma_y)$, where $\sigma_j$ are the
Pauli matrices. The electrostatic potential energy $V(r)$ depends only on
distance from the origin in polar coordinates $(r,\varphi)$ and is given by
\begin{equation}
  \label{v0mpot}
  V(r) = -V_0 \times
  \begin{cases}
    \,\frac{2^m|r-R_{\rm av}|^m}{|R_{\rm o}-R_{\rm i}|^m}
    &  \text{if }\ |r\!-\!R_{\rm av}| \leqslant \frac{R_{\rm o}-R_{\rm i}}{2}, \\
    \,1  &  \text{if }\ |r\!-\!R_{\rm av}| > \frac{R_{\rm o}-R_{\rm i}}{2}, 
  \end{cases}
\end{equation}
where we have defined $R_{\rm av}=(R_{\rm i}+R_{\rm o})/2$ with $R_{\rm i}$
and $R_{\rm o}$ being the inner and outer radii of the disk.
We also note that the limit of $V_0\rightarrow\infty$ and
$m\rightarrow\infty$ restores the familiar rectangular barrier of
an~infinite height \cite{Ryc09,Ryc10}. 

Because of the symmetry, the wave function can be written in a~form
\begin{equation}
\Psi_j(r,\varphi) = e^{i(j-1/2)\varphi}\left(
    \begin{array}{c} \chi_a \\ \chi_be^{i\varphi} \\ \end{array}
  \right),
\end{equation}
where $j=\pm{1/2},\pm{3/2},\dots$ is the total angular-momentum
quantum number. In the leads, $r<R_{\rm i}$ or $r>R_{\rm o}$, the electrostatic
potential energy is constant, $V(r)=-V_0$.
In the case of electron doping, $E>-V_0$, the spinors $(\chi_a,\chi_b)^T$
for the incoming (i.e., propagating from $r=0$) and outgoing
(i.e., propagating from $r=\infty$) waves are given, up to the normalization,
by
\begin{equation}
  \label{chijleads}
  \chi_j^{\rm in} \!= \left(
  \begin{array}{c}
    H_{j-1/2}^{(2)}(kr) \\
    iH_{j+1/2}^{(2)}(kr) \\
  \end{array}
  \right),
  \ \ \ 
  \chi_j^{\rm out} \!= \left(
  \begin{array}{c}
    H_{j-1/2}^{(1)}(kr) \\
    iH_{j+1/2}^{(1)}(kr) \\
  \end{array}
  \right),
  \end{equation}
where $H_\nu^{(1)}(\rho)$ [$H_\nu^{(2)}(\rho)$] is the Hankel function of the
first [second] kind, $k=|E+V_0|/(\hbar{}v_F)$, and we have set $B=0$ in the
leads \cite{hankfoo}.
(The wavefunctions for $B\neq{}0$ are given explicitly in
Refs.\ \cite{Che06,Ryc10}.)
Full wavefunctions in the leads, for a~given $j$, can be written as 
\begin{align}
  \chi_j^{({\rm i})} &=
  \chi_j^{\rm in} + r_j\chi_j^{\rm out},
    \  &r<R_{\rm i},
  \label{chij1hank} \\
  \chi_j^{({\rm o})} &= t_j\chi_j^{\rm in}, 
    \  &r>R_{\rm o}, \label{chij2hank} 
\end{align}
with the reflection (and transmission) amplitudes $r_j$ (and $t_j$). 

For the disk area, $R_{\rm i}<r<R_{\rm o}$, we have $B\neq{}0$ and the
position-dependent $V(r)$. Eq.\ (\ref{direqvr}) brought us to
the system of ordinary differential equations
for spinor components 
\begin{align}
  \chi_a' &= \left(\frac{j-1/2}{r}+\frac{eBr}{2\hbar}\right)\chi_a
  +i\,\frac{E-V(r)}{\hbar{}v_F}\chi_b,
  \label{phapri} \\
  \chi_b' &= i\,\frac{E-V(r)}{\hbar{}v_F}\chi_a
  -\left(\frac{j+1/2}{r}+\frac{eBr}{2\hbar}\right)\chi_b,
  \label{phbpri} 
\end{align}
which has to be integrated numerically for all $j$-s. 
To reduce round-off errors that occur in finite-precision arithmetic due to
exponentially growing (or decaying) spinor components, we have divided the
interval $R_{\rm i}\dots{}R_{\rm o}$ into $M$ equally wide parts, bounded
by $R_{\rm i}^{(l)}<r<R_{\rm o}^{(l)}$, with $l=0,1,\dots,M-1$, and
\begin{equation}
  R_{\rm i}^{(l)}=R_{\rm i}+l\,\frac{R_{\rm o}-R_{\rm i}}{M}, \ \ \ \ \ 
  R_{\rm o}^{(l)}=R_{\rm i}^{(l+1)}. 
\end{equation}
The resulting wave function for the $l$-th interval then has the form 
\begin{equation}
\label{chijdisknum}
  \chi_j^{(l)}=A_j^{(l)}\chi_j^{(l),{\rm I}}+B_j^{(l)}\chi_j^{(l),{\rm II}}, 
\end{equation}
where $\chi_j^{(l),{\rm I}}$, $\chi_j^{(l),{\rm II}}$ denote the two linearly
independent solutions that we obtained numerically by solving
Eqs.\ (\ref{phapri},\ref{phbpri}) with two different initial conditions, 
$\left.\chi_j^{(l),{\rm I}}\right|_{r=R_{\rm i}^{(l)}}=(1,0)^T$ and  
$\left.\chi_j^{(l),{\rm II}}\right|_{r=R_{\rm i}^{(l)}}=(0,1)^T$. 
$A_j^{(l)}$ and $B_j^{(l)}$ are arbitrary complex coefficients. 

The matching conditions, namely
\begin{align}
  \chi_j^{({\rm i})}(R_{\rm i}) &= \chi_j^{(0)}(R_{\rm i}),
    \label{match-i} \\
  \chi_j^{(l)}(R_{\rm o}^{(l)}) &= \chi_j^{(l+1)}(R_{\rm i}^{(l+1)}),
  \ \ \ \ 
  l = 0,\dots,M-2,
  \label{match-l} \\
  \chi_j^{(M-1)}(R_{\rm o}) &= \chi_j^{({\rm o})}(R_{\rm o}),
  \label{match-o} 
\end{align}
are equivalent to the system of $2M+2$ linear equations for the unknowns 
$A_j^{(0)}$, $B_j^{(0)}$, \dots, $A_j^{(M-1)}$, $B_j^{(M-1)}$, $r_j$, and $t_j$,
which can be written as
\begin{widetext}
$$
  \left[
  \begin{matrix}
    -\chi_{j,a}^{\rm out}(R_{\rm i})
    & \chi_{j,a}^{(0),{\rm I}}(R_{\rm i}^{(0)})
    & \chi_{j,a}^{(0),{\rm II}}(R_{\rm i}^{(0)})
    & & & &  \\
    -\chi_{j,b}^{\rm out}(R_{\rm i})
    & \chi_{j,b}^{(0),{\rm I}}(R_{\rm i}^{(0)})
    & \chi_{j,b}^{(0),{\rm II}}(R_{\rm i}^{(0)})
    & & & &  \\
    0 &  \chi_a^{(0),{\rm I}}(R_{\rm o}^{(0)})
    & \chi_{j,a}^{(0),{\rm II}}(R_{\rm o}^{(0)})
    &  & & &  \\
    0 &  \chi_b^{(0),{\rm I}}(R_{\rm o}^{(0)})
    & \chi_{j,b}^{(0),{\rm II}}(R_{\rm o}^{(0)})
    & \ddots & & &  \\
    &  &  & \ddots &
    \chi_{j,a}^{(\overline{M}),{\rm I}}(R_{\rm i}^{(\overline{M})}) 
    & \chi_{j,a}^{(\overline{M}),{\rm II}}(R_{\rm i}^{(\overline{M})})
    & 0 \\
    &   &  &  &
    \chi_{j,b}^{(\overline{M}),{\rm I}}(R_{\rm i}^{(\overline{M})}) 
    & \chi_{j,b}^{(\overline{M}),{\rm II}}(R_{\rm i}^{(\overline{M})})
    & 0 \\
    &  &  &  &
    \chi_a^{(\overline{M}),{\rm I}}(R_{\rm o}^{(\overline{M})}) &
    \chi_{j,a}^{(\overline{M}),{\rm II}}(R_{\rm o}^{(\overline{M})})
    & -\chi_{j,a}^{\rm in}(R_{\rm o}) \\
    &  &  &  &
    \chi_b^{(\overline{M}),{\rm I}}(R_{\rm o}^{(\overline{M})})
    & \chi_{j,b}^{(\overline{M}),{\rm II}}(R_{\rm o}^{(\overline{M})})
    & -\chi_{j,b}^{\rm in}(R_{\rm o}) \\
  \end{matrix}
  \right]
$$
\begin{equation}
  \label{lsysABrt}
  \times\ 
  \left[
    \begin{matrix}
      r_j \\ A_j^{(0)} \\ B_j^{(0)} \\ \vdots
      \\ A_j^{(M-1)} \\ B_j^{(M-1)} \\ t_j \\
    \end{matrix}
  \right]
  =
  \left[
    \begin{matrix}
      \,\chi_{j,a}^{\rm in}(R_{\rm i})\, \\
      \,\chi_{j,b}^{\rm in}(R_{\rm i})\, \\
      0 \\
      \vdots \\
      0 \\
    \end{matrix}
  \right], 
\end{equation}
\end{widetext}
where we have explicitly written the spinor components of the relevant
wavefunctions appearing in Eqs.\ (\ref{match-i}),
(\ref{match-l}), and (\ref{match-o}) and defined $\overline{M}=M-1$. 

Analogously, assuming the scattering from $r=\infty$, we replace Eqs.\
(\ref{chij1hank},\ref{chij2hank}) with
\begin{align}
  \chi_j^{({\rm i})} &= t_j'\chi_j^{\rm out},
    \  &r<R_{\rm i},
  \label{chij1hankpr} \\
  \chi_j^{({\rm o})} &= \chi_j^{\rm out} + r_j'\chi_j^{\rm in}, 
    \  &r>R_{\rm o}, \label{chij2hankpr} 
\end{align}
and follow the consequtive steps mentioned above to obtain the linear system 
\begin{equation}
  {\mathbb{A}}_j\times
  \left[
    \begin{matrix}
      t_j' \\ {A_j^{(0)}}' \\ {B_j^{(0)}}' \\ \vdots
      \\ {A_j^{(M-1)}}' \\ {B_j^{(M-1)}}' \\ r_j' \\
    \end{matrix}
  \right]
  =
  \left[
    \begin{matrix}
      0 \\
      \vdots \\
      0 \\
      \,\chi_{j,a}^{\rm out}(R_{\rm o})\, \\
      \,\chi_{j,b}^{\rm out}(R_{\rm o})\, \\
    \end{matrix}
  \right], 
\end{equation}
where $\mathbb{A}_j$ is the main matrix as in Eq.\ (\ref{lsysABrt})
and $r_j'$ ($t_j'$) being the reflection (transmission) amplitude for the
scattering from the outer lead. 
(Note that the elements of the $\mathbb{A}_j$ matrix are unchanged; hence, 
Eqs.\ (\ref{phapri},\ref{phbpri}) need only to be integrated numerically 
once for a~given $j$.) The scattering matrix, 
\begin{equation}
  {\mathbb{S}}_j=
  \left[
    \begin{matrix}
      r  &  t' \\
      t  &  r' \\
    \end{matrix}
  \right], 
\end{equation}
contains all the amplitudes mentioned above. Conservation of electric xharge
implies the unitarity of the ${\mathbb{S}}_j$ matrix,
${\mathbb{S}}_j{\mathbb{S}}_j^\dagger = {\mathbb{S}}_j^\dagger{\mathbb{S}}_j
= {\mathbb{I}}$ (where ${\mathbb{I}}$ is the identity matrix).
The deviation from unitarity due to numerical errors, i.e.,
 $\mbox{max}_j\left(|\epsilon_j|\right)$ with
 $\epsilon_j=\mbox{Tr}\left({\mathbb{S}}_j^\dagger{\mathbb{S}}_j-{\mathbb{I}}
 \right)$, provides a~useful measure of the computational accuracy. 

Since linear systems for different $j$-s are decoupled, numerous software
packages can be used to find their solutions up to machine precision.
We use the double precision LAPACK routine {\tt zgesv}, see Ref.\ 
\cite{zgesv99}.
The transmission probabilities are calculated as $T_j=|t_j|^2$. 

For heavily-doped leads ($V_0\rightarrow\infty$) the wave functions
given by Eq.\ (\ref{chijleads}) simplify to 
\begin{equation}
  \label{chijasym} 
  \chi_j^{({\rm in})} =
    \frac{e^{iKr}}{\sqrt{r}}
    \left(\begin{array}{c} 1 \\ 1 \\ \end{array}\right), 
    \ \ \ \ 
    \chi_j^{({\rm out})} =
    \frac{e^{-iKr}}{\sqrt{r}}
    \left(\begin{array}{c} 1 \\ -1 \\ \end{array}\right),
\end{equation}
with $K=|E+{}V_0|/(\hbar{}v_F)\rightarrow{}\infty$.
In particular, for $m\rightarrow\infty$ and $M=1$, closed-form expressions
for $T_j$-s were found, either for $B=0$ \cite{Ryc09} or for $B\neq{}0$
\cite{Ryc10}.
However, we have found that the available implementations of hypergeometric
functions that occur in these expressions lead to numerical
stability problems when calculating thermoelectric properties in the quantum
Hall regime.
Therefore, a~procedure described in this subsection is applied directly
in all the following numerical examples, with wave functions
in the leads given by Eqs.\ (\ref{chij1hank},\ref{chij2hank}) for
$m<+\infty$ (smooth potential barriers). 
For $m\rightarrow\infty$ (rectangular barrier), both the finite 
and infinite doping in the leads are studied for comparison.

The numerical integration of Eqs.\ (\ref{phapri},\ref{phbpri}) was performed
for each of $M=20$ intervals using a~standard fourth-order Runge-Kutta
(RK4) algorithm.
For $R_{\rm o}=2R_{\rm i}=1000\,$nm and $B<0.5\,$T, 
a~spatial step of $0.5\,$pm was sufficient to reduce 
the unitarity error down to $\mbox{max}\left(|\epsilon_j|\right)<10^{-8}$. 
The summation over the modes was stopped when $T_j<10^{-7}$.
The Dormand-Prince method \cite{simfoo} was also implemented; however,
no significant differences in the transmission probabilities
were found compared to RK4.

\subsection{Thermoelectric characteristics}

Within the Landauer-B\"{u}ttiker formalism, the linear-response conductance 
\cite{Lan57,But85} and other thermoelectric properties \cite{Pau03,Esf06}
can be calculated from the transmission-energy dependence 
\begin{equation}
  \label{teland}
  {\cal T}(E) = \sum_{j=-j_{\rm max}}^{j_{\rm max}}{}T_j(E), 
\end{equation}
where $j_{\rm max}=\lfloor{}KR_{\rm i}\rfloor-\frac{1}{2}$ (for heavily-doped
leads, $j_{\rm max}\rightarrow{}\infty$), via dimensionless integrals 
\begin{equation}
  \label{llnmom}
  {L}_n= (k_BT)^{-n}\int{}dE\,{\cal T}(E)\left(
  -\frac{\partial{}f_{\rm FD}}{\partial{}E}\right)(E-\mu)^n,
\end{equation}
with $f_{\rm FD}(\mu,T,E)=
1/\left[\,\exp\left((E\!-\!\mu)/k_BT\right)+1\,\right]$ 
the Fermi-Dirac distribution function and the chemical potential $\mu$. 
In particular, 
\begin{align}
  G &= \frac{g_sg_ve^2}{h}L_0 \ \ \ \ (\text{the conductance}),
  \label{gland} \\
  S &= \left.\frac{dU}{dT}\right|_{I=0} = \frac{k_BL_1}{eL_0}
  \ \ \ \ (\text{the Seebeck coefficient}),
  \label{sland} \\
  {\cal L} &= \frac{K_{\rm el}}{TG} = \frac{k_B^2(L_0L_2\!-\!L_1^2)}{e^2L_0^2}
  \ \ \ (\text{the Lorentz number}),
  \label{llndef}
\end{align}
where $g_s=g_v=2$ are spin and valley degeneracies, $dU/dT|_{I=0}$ is the
voltage derivative with respect to the temperature difference between the
leads at zero electric current, and $K_{\rm el}$ is the electronic part of
the thermal conductance. 

For zero temperature, Eq.\ (\ref{gland}) reduces to
\begin{equation}
  \label{gtef}
  G(T\!\rightarrow{}\!0) = g_0{\cal T}(E_F), 
\end{equation}
where we have defined the conductance quantum $g_0=4e^2/h$ and the Fermi
energy $E_F$ ($=\mu$ for $T=0$).
In turn, the zero-temperature conductance provides a~direct insight into
the transmission-energy dependence. 

For some specific ${\cal T}(E)$, integrals in Eqs.\ (\ref{sland},\ref{llndef})
can be calculated analytically. In particular, ${\cal T}(E)\approx{}$const
leads to $S\approx{}0$ and ${\cal L}\approx{}{\cal L}_0=(\pi^2/3)\,k_B^2/e^2$,
which defines the Wiedemann-Franz law for metals \cite{Kit05}.
For gapless Dirac systems, the corresponding approximation is
${\cal T}(E)\approx{}{\cal C}|E|$, with a~constant ${\cal C}>0$, for which
both $S$ and ${\cal L}$ can be expressed by the polylogarithm function
of $\mu/k_BT$ \cite{Sus18}, with a~universal (${\cal C}$-independent) maxima
$S_{\rm max}\simeq{}1.0023\,k_B/e$ and
${\cal L}_{\rm max}\simeq{}2.3721\,{\cal L}_0$.
The latter value was first reported in the context of $d$-wave systems
\cite{Sha03}, before being found again for Dirac materials
\cite{Sai07,Yos15,Ing15}. 

In the presence of a~transport gap, one can consider a~simplified model for
${\cal T}(E)$, given by
\begin{equation}
  \label{temodel}
  {\cal T}_{\rm model}(E) =
  {\cal A}\delta(E) + {\cal B}\delta(E-\Delta{}E),
\end{equation}
where ${\cal A}>0$, ${\cal B}>0$ are the constants, and $\delta(x)$
is the Dirac delta function. 
Generalizing the derivations presented in Refs.\ \cite{Sus19} and
\cite{Ryc21a} to the asymmetric case (${\cal A}\neq{\cal B}$), one
finds easily 
\begin{align}
  \frac{S_{\rm max}\!-\!S_{\rm min}}{2} &\simeq{}
  \frac{k_B}{e}\!\left[\sqrt{u(u\!-\!1)} - \ln(\sqrt{u}+\sqrt{u\!-\!1})\right],
  \label{smaxsmin}
  \\
  {\cal L}_{\rm max} &\simeq{} \left(\frac{k_B}{e}\right)^2u^2 = 
  \left(\frac{\Delta{}E}{2eT}\right)^2,
  \label{llnmax}
\end{align}
where $u=\Delta{}E/(2k_BT)$  and the asymptotic equalities correspond to
$u\gg{}1$. The remaining symbols in Eq.\ (\ref{smaxsmin}) are the maximum
($S_{\rm max}$) and the minimum ($S_{\rm min}$) Seebeck coefficient in the
interval of $0<\mu<\Delta{E}$.
Note that the right-hand sides in Eqs.\ (\ref{smaxsmin}) and (\ref{llnmax})
depend only on $u$ and the fundamental constants. 

In a~case where more $\delta$-shaped peaks appear in the transmission spectrum
${\cal T}(E)$, as might be expected for the quantum Hall regime, the
approximations given by Eqs.\ (\ref{smaxsmin}) and (\ref{llnmax})
are also valid, provided that a~gap is identified with the maximum 
interval between the peaks ($\Delta{}E=\Delta{}E_{\rm max}$).
The monotonicity
of the right-hand sides in Eqs.\ (\ref{smaxsmin}) and (\ref{llnmax})
guarantees that the resulting approximations, for $\Delta{}E=\Delta{}E_{\rm max}$,
correspond to the global maxima of the relevant quantities (i.e., the
thermopower amplitude and the Lorentz number) as functions of $\mu$.

\begin{figure*}[!t]
  \includegraphics[width=\linewidth]{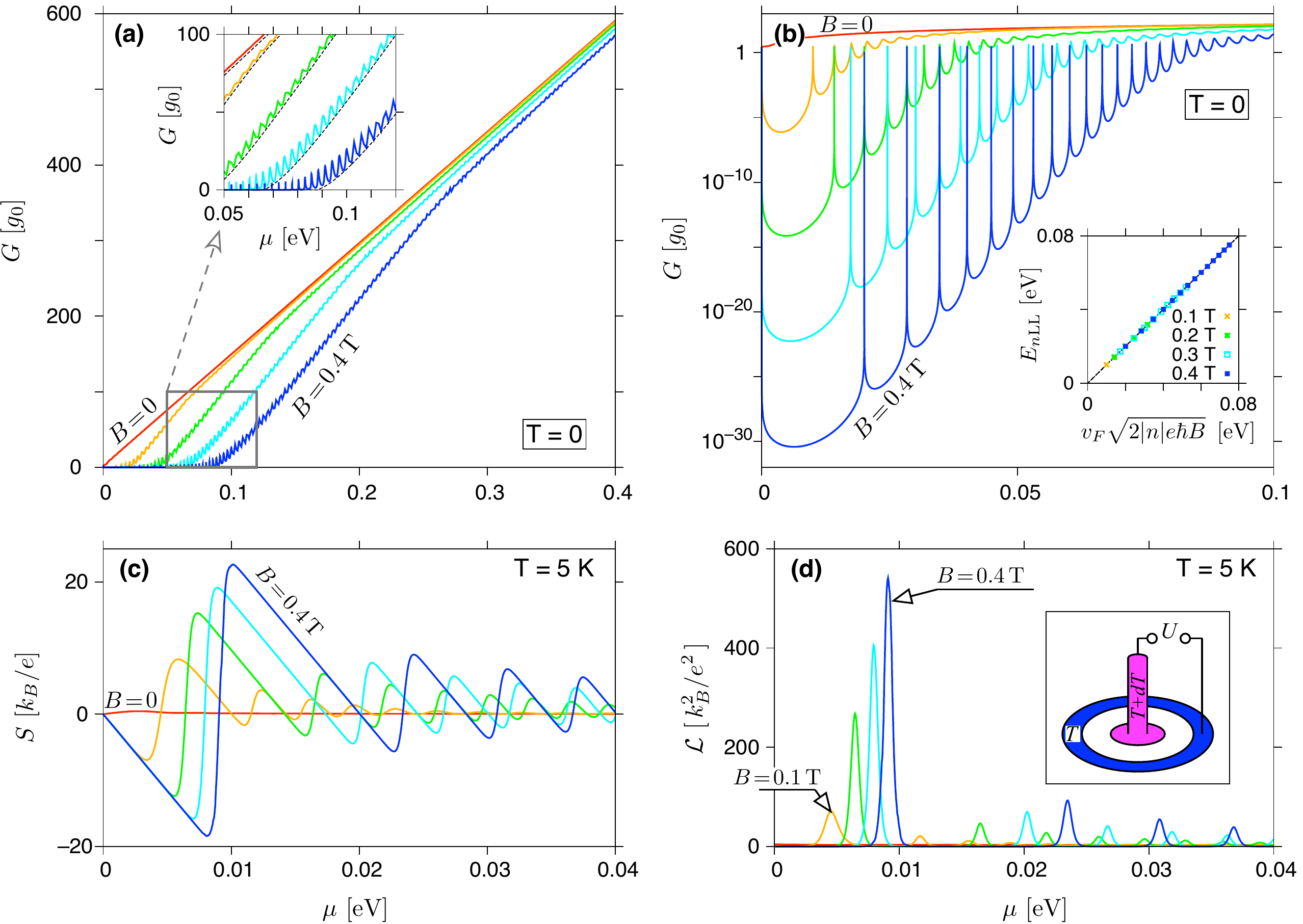}
  \caption{ \label{gslmax4pan:fig}
    (a,b) Zero-temperature conductance, (c) the Seebeck coefficient and (d)
    the Lorentz number, both for $T=5\,$K,  
    for the system of Fig.\ \ref{setup3:fig} with
    $R_{\rm o}=2R_{\rm i}=1000\,$nm and the rectangular potential barrier
    [$V_0,m\rightarrow{}\infty$; see Eq.\ (\ref{v0mpot})] displayed
    as functions of the chemical potential.
    The magnetic field is varied from $B=0$ (red solid lines in all plots)
    to $B=0.4\,$T (blue solid lines) with the steps of $0.1\,$T.
    Inset in (a) is a~zoom-in, with black dashed lines depicting the
    incoherent conductance (see Appendix~A). (b) shows same data as (a), but
    using the semilogarithmic scale, with the inset presenting positions of
    the actual transmission maxima for $B>0$ ($E_{n{\rm LL}}$) versus the
    values for bulk graphene [see Eq.\ (\ref{enllbulk})].
    A~setup for thermoelectric measurements is also depicted
    [see inset in (d)]. 
  }
\end{figure*}

\begin{figure*}[!t]
  \includegraphics[width=\linewidth]{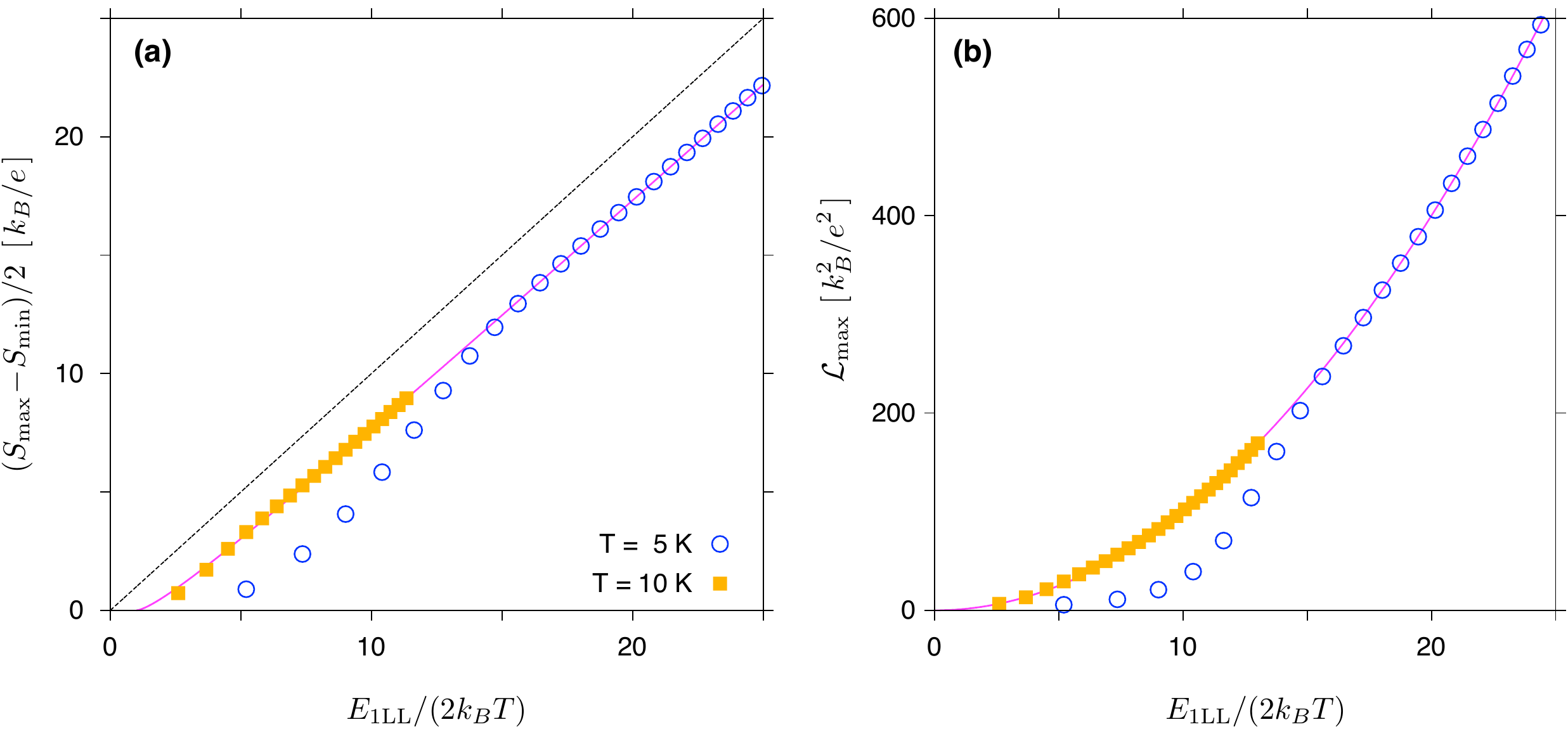}
  \caption{ \label{smaxlmax2pan:fig}
    (a) Maximum amplitude of the Seebeck coefficient and (b) maximum
    Lorentz number for same system as in Fig.\ \ref{gslmax4pan:fig} at
    $T=5\,K$ (open symbols) and $T=10\,$K (closed symbols) as functions
    of the maximum interval between Landau levels 
    of $\Delta{}E_{\rm max}\equiv{}E_{1{\rm LL}}=|E_{0{\rm LL}}-E_{\pm{}1{\rm LL}}|$.
    Solid lines depict the asymptotic expressions given in Eqs.\
    (\ref{smaxsmin}) and (\ref{llnmax}).
    Dashed line in (a) corresponds to the Goldsmid-Sharp relation,
    see Eq.\ (\ref{smaxgs}), with $E_g=E_{1{\rm LL}}$. 
  }
\end{figure*}

\section{Results and discussion}
\label{resdis}

\subsection{Zero-temperature conductance}

Before discussing the thermoelectric properties, we present
zero-temperature conductance spectra, wjich are related to
the transmission-energy dependence via Eq.\ (\ref{gtef}),
and thus represent the input data for the calculatoin of the Seebeck
coefficient and the Lorentz number.
For the rectangular barrier of infinite height, 
$m,V_0\rightarrow{}\infty$ in Eq.\ (\ref{v0mpot}), the spectra
are particle-hole symmetric, and it is sufficient to consider 
$\mu\geqslant{}0$. 

In Figs.\ \ref{gslmax4pan:fig}(a) and \ref{gslmax4pan:fig}(b) we display
the disk conductance as a~function of $\mu=E_F$ for $T=0$.
Resonances via Landau levels, centered near energies very close to the
corresponding values for bulk graphene, 
\begin{equation}
\label{enllbulk}
  E_{n{\rm LL}}\approx{}E_{n{\rm LL}}^{{\rm bulk}}=
  \mbox{sgn}(n)v_F\sqrt{2|n|e\hbar{}B}, 
\end{equation}
where $n$ is an integer (without loss of generality, we hereinafter
suppose $B>0$),
are clearly visible starting from a~moderate value of $B=0.1\,$T.
More generally, one can expect the resonances to be visible up to the $n$-th 
one if
\begin{equation}
  \label{enllec1}
  E_{n{\rm LL}}\lesssim{}
  E_{c,1}=\frac{v_FeB}{2}\left(R_{\rm o}-R_{\rm i}\right), 
\end{equation}
where $E_{c,1}$ denotes the threshold energy, below which the cyclotron
diameter $2r_c=2|E_F|/(v_FeB)<R_{\rm o}-R_{\rm i}$ and the incoherent
transmission vanishes (see Appendix~A).
For instance, for $R_{\rm o}=2R_{\rm i}=1000\,$nm discussed throughout
the paper, Eq.\ (\ref{enllec1}) gives
\begin{equation}
|n|\lesssim{}47.47\times{}B\,[\text{T}], 
\end{equation}
coinciding with the number of well-separated maxima visualized in
semi-logarithmic scale in Fig.\ \ref{gslmax4pan:fig}(b).  

It is also visible, for $|E_F|>E_{c,1}$, that the actual $G$ grows rapidly
with increasing $|E_F|$,  closely following the prediction for incoherent
transport [see Fig.\ \ref{gslmax4pan:fig}(a) and the inset].
This observation leads to the question whether $E_{1LL}$, or rather $2E_{c,1}$, 
defines the relevant transport gap to be substituted into Eqs.\
(\ref{smaxsmin}) and (\ref{llnmax})?

This problem is solved via the numerical analysis of $S$ and ${\cal L}$
which will be presented next.

\subsection{Thermopower and the Lorentz number}

The Seebeck coefficient and the Lorentz number, calculated from
Eqs.\ (\ref{sland}) and (\ref{llndef}) for $T=5\,$K, are shown 
in Figs.\ \ref{gslmax4pan:fig}(c) and \ref{gslmax4pan:fig}(d) as functions
of the chemical potential.
Again, the particle-hole symmetry of ${\cal T}(E)$ guarantees that $S(\mu)$
is odd and ${\cal L}(\mu)$ is even on $\mu\rightarrow{}-\mu$, and
it is sufficient to consider $\mu\geqslant{}0$. 

In the quantum Hall regime, i.e., for $|\mu|<E_{c,1}$, see Eq.\
(\ref{enllec1}), the ${\cal T}(E)$ function
consists of narrow peaks, each centered at $E_{n{\rm LL}}$ (see previous
subsection).
In such a~case, reliable numerical estimations of the integrals $L_0$, $L_1$,
and $L_2$ [see Eq.\ (\ref{llnmom})] requires sufficiently dense sampling
of ${\cal T}(E)$ near $E\approx{}E_{n{\rm LL}}$. 

The analytic structure of 
Eqs.\ (\ref{sland}) and (\ref{llndef}) results in the following features
visible in Figs.\ \ref{gslmax4pan:fig}(c) and \ref{gslmax4pan:fig}(d):
First, each of the consecutive intervals, i.e., $E_{0{\rm LL}}<E<E_{1{\rm LL}}$,
$E_{1{\rm LL}}<E<E_{2{\rm LL}}$, etc., contains a~local minimum and a~local
maximum of $S$, surrounding an odd zero of $S$ (even zeros occur for the
resonances at $\mu\approx{}E_{n{\rm LL}}$), 
which corresponds to a~local maximum of ${\cal L}$. 
Second, the global extrema (corresponding to $S_{\rm min}$, $S_{\rm max}$,
or ${\cal L}_{\rm max}$) are all in the first interval,
$E_{0{\rm LL}}<E<E_{1{\rm LL}}$, characterized by the maximum width
($\Delta{}E_{\rm max}\equiv{}E_{1{\rm LL}}-E_{0{\rm LL}}=E_{1{\rm LL}}$). 

In Fig.\ \ref{smaxlmax2pan:fig} we plot the values of
$(S_{\rm max}-S_{\rm min})/2$ and ${\cal L}_{\rm max}$ against the dimensionless
variable $E_{1{\rm LL}}/(2k_B{}T)$, for the two values of $T=5\,$K and $10\,$K.
Results of the numerical integration and subsequent optimization with respect
to the chemical potential $\mu$ [datapoints] closely follow the approximations
given in Eqs.\ (\ref{smaxsmin},\ref{llnmax}) [solid lines], starting from
$E_{1{\rm LL}}/(2k_B{}T)\gtrsim{}10$.
The Goldsmid-Sharp formula [dashed line] produces a~noticeable offset when
compared to the Landauer-B\"{u}ttiker results, but can still be used as
a~less accurate approximation for $(S_{\rm max}-S_{\rm min})/2$ for large
$E_{1{\rm LL}}/(2k_B{}T)$. 

These results support our conjecture that the model ${\cal T}(E)$, given
by Eq.\ (\ref{temodel}), is able to reproduce the basic thermoelectric
properties of graphene disk in the quantum Hall regime.
Although it may seem surprising, at least at first glance, that the model
${\cal T}(E)$ with $\Delta{}E=E_{1{\rm LL}}$ reproduces the actual numerical
results, whereas the energy scale of $2E_{c,1}\gg{}E_{1{\rm LL}}$ [see Eq.\
(\ref{enllec1})] seems to be irrelevant. 
However, for thermal excitation energies 
$k_B{}T\ll{}E_{1{\rm LL}}\ll{}2E_{c,1}$, the detailed behavior of the actual
${\cal T}(E)$ [see Eq.\ (\ref{teland})]
for $E-E_{0{\rm LL}}=E\lesssim{}-k_BT$ or $E-E_{1{\rm LL}}\gtrsim{}k_B{}T$
does not affect the integrals $L_0$, $L_1$, $L_2$ 
(note that the full width at half maximum
for $-\partial{}f_{\rm FD}/\partial{}E$ in Eq.\ (\ref{llnmom}) is
$\approx{}\!3.53\,k_B{}T\,$) when $0<\mu<E_{1{\rm LL}}$). 
For this reason, a~model with  $\Delta{}E=E_{1{\rm LL}}$ captures 
the essential features of the actual  ${\cal T}(E)$, while focussing
on the thermoelectric properties considered here.

\begin{figure*}[!t]
  \includegraphics[width=\linewidth]{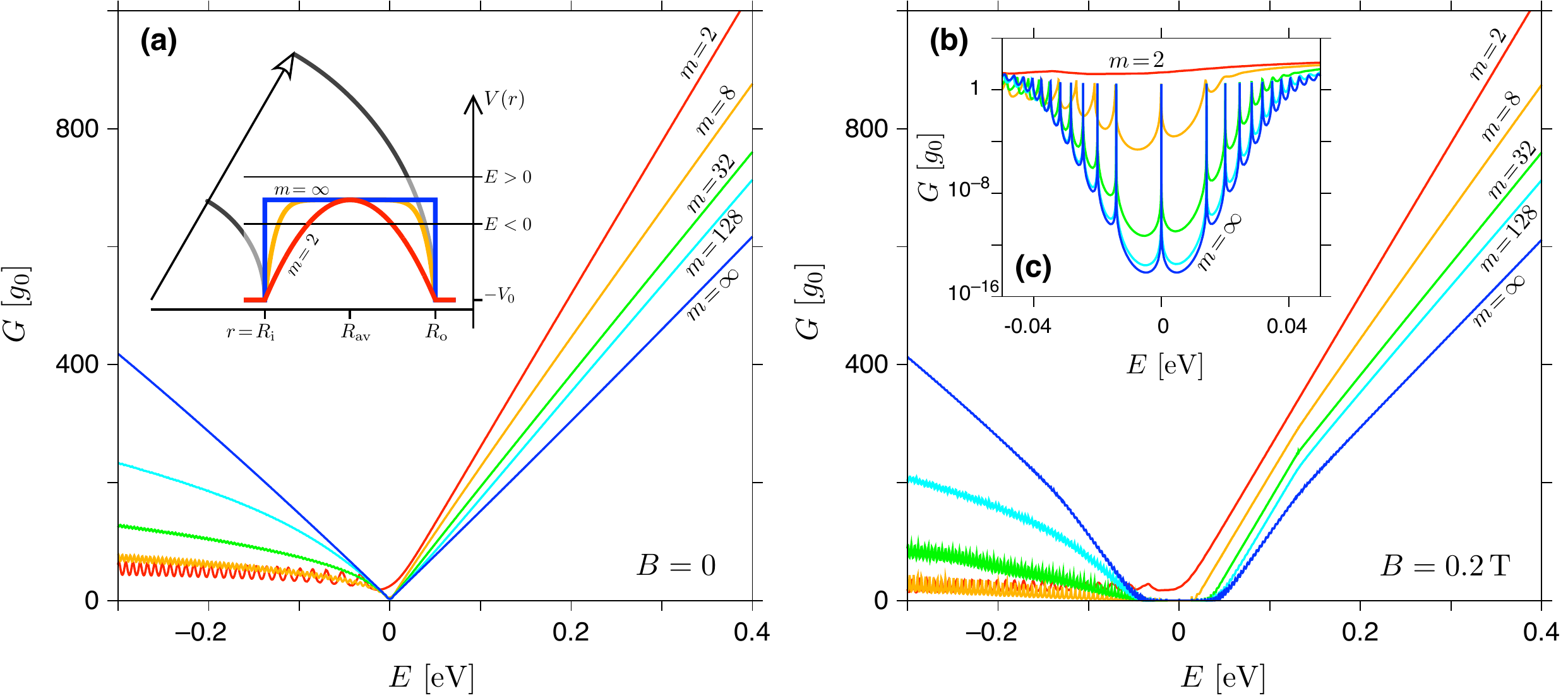}
  \caption{ \label{gmpots2pan:fig}
    Zero-temperature conductance at (a) $B=0$ and (b) $B=0.2\,$T versus
    the Fermi energy. The disk radii are same as Fig.\ \ref{gslmax4pan:fig}, 
    the barrier height [see Eq.\ (\ref{v0mpot})] is fixed at
    $V_0=t_0/2=1.35\,$eV, the parameter $m$ is specified for each line.
    Inset in (a) shows selected potential profiles.
    (c) Zoom-in, for low energies, with same datasets as in (b) displayed
    in semi-logarithmic scale. 
  }
\end{figure*}

\subsection{Smooth potential barriers}
For the sake of completeness, in this section we also revisit the effects
of smooth potential barriers, considered earlier for zero magnetic field
\cite{Ryc22}.
The electrostatic potential energy is given by Eq.\ (\ref{v0mpot}), where
the barrier height is fixed at $V_0=t_0/2=1.35\,$eV (for selected profiles
for $m=2$, $8$, and $\infty$, see Fig.\ \ref{gmpots2pan:fig}) and the radii
$R_{\rm o}=2R_{\rm i}=1000\,$nm again. 

For a~finite barrier height, the particle-hole symmetry of the conductance
spectrum $G(E)$ is absent, even for the rectangular barrier ($m=\infty$).
However, for sufficiently low energies $|E|\ll{}V_0$ and large $m$, the
Fermi wavelength $\lambda_F=hv_F/|E|$ becomes longer than a~characteristic
length scale of a potential jump, i.e., $\Delta{}r=(L-L_{\rm eff})/2$ where
the sample length is $L=R_{\rm o}-R_{\rm i}=500\,$nm, and
\begin{equation}
  \label{ldiffm}
  {L_{\rm eff}^{(m)}} = L\left(\frac{\hbar{}v_F}{LV_0}\right)^{1/m}   
\end{equation}
is the so-called diffusive length defined via
$V(R_{\rm av}\pm{}L_{\rm eff}/2)=-\hbar{}v_F/L$ (see Ref.\ \cite{Ryc21b}).
The value of $\hbar{}v_F/L$ can be attributed to the energy uncertainty
corresponding to a~typical time of flight $\sim{}L/v_F$ (up to the order
of magnitude). We further note that $L_{\rm eff}^{(m)}\rightarrow{}L$ for
$m\rightarrow{}\infty$. 
If $\lambda_F\gg{}\Delta{}r$, the potential profile can be considered as
approximately flat, and the approximate symmetry upon $E\rightarrow{}-E$
can be observed for the corresponding zero-field spectra shown in
Fig.\ \ref{gmpots2pan:fig}(a). 

For $B=0.2\,$T, see Figs.\ \ref{gmpots2pan:fig}(b) and
\ref{gmpots2pan:fig}(c), the approximate symmetry is also visible. 
In addition, it is worth noting that for $\lambda_F\gg{}\Delta{}r$
the lowest LLs are well pronounced, and their positions are almost
unaffected compared to the infinite-barrier case [see previous subsection,
Fig.\ \ref{gslmax4pan:fig}(b)]. 

In both cases, i.e., for $B=0$ and $B=0.2\,$T, the presence of two circular
p-n junctions for $E<0$ leads to a~suppressed  conductance compared to
$E>0$, with well-pronounced conductance oscillations due to quasi-bounded
states (especially for smaller $m$).
Due to such an asymmetry, the global conductance minimum $G_{\rm min}$ in the
quantum Hall regime is typically reached
in the energy interval of $-E_{1{\rm LL}}<E<0$, with the exception of
the parabolic profile ($m=2$),  for which resonances with LLs are
obliterated.

\begin{figure*}[!t]
  \includegraphics[width=\linewidth]{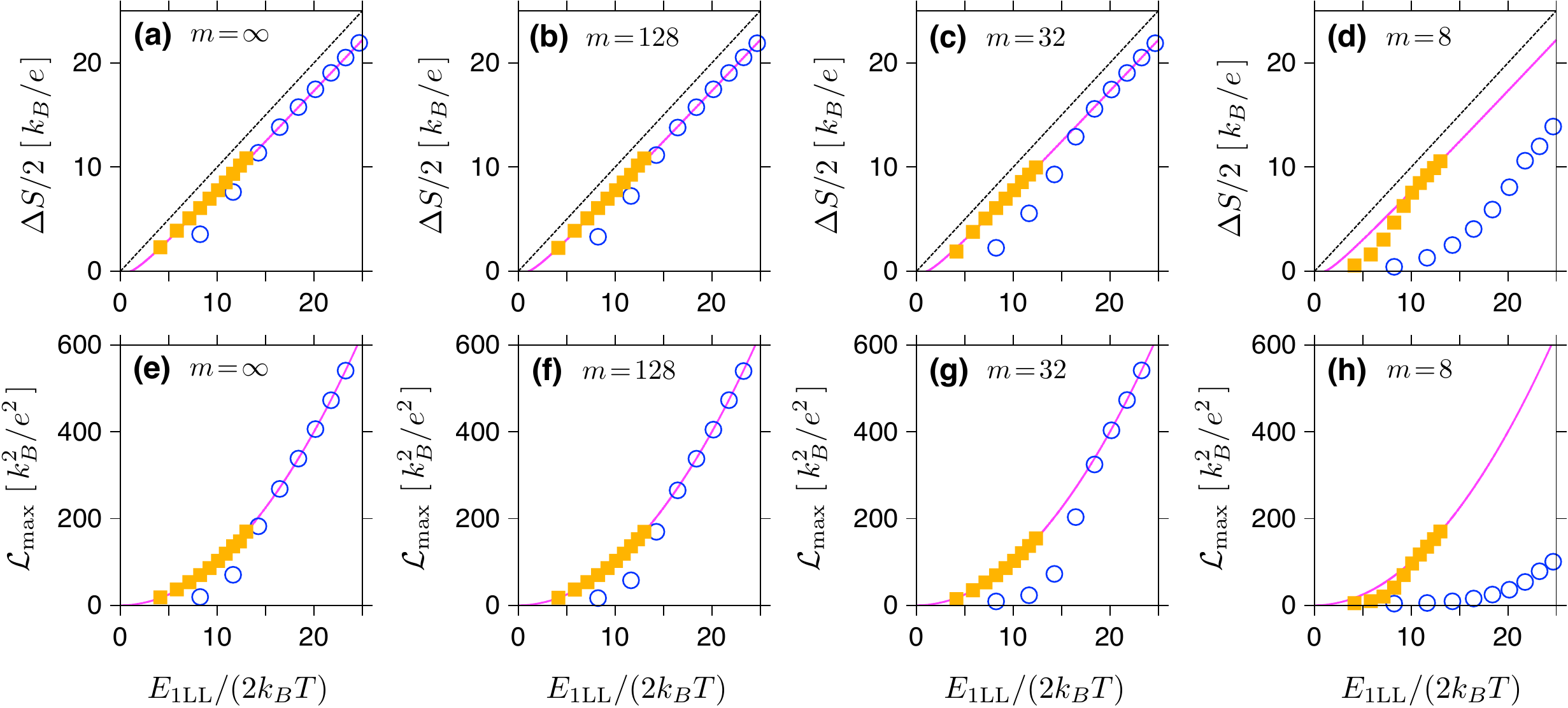}
  \caption{ \label{slmpots8pan:fig}
    (a)--(d) Maximum amplitude of the Seebeck coefficient
    $\Delta{}S=(S_{\rm max}-S_{\rm min})/2$ and
    (e)--(h) maximum Lorentz number for same system as in Fig.\
    \ref{gmpots2pan:fig} and selected values of the exponent $m$ defining
    the potential profile [see Eq.\ (\ref{v0mpot})] displayed as the bulk
    Landau-level energy $E_{1{\rm LL}}$ obtained from Eq.\ (\ref{enllbulk})
    with $n=1$.
    The line/color encoding is same as in Fig.\ \ref{smaxlmax2pan:fig}. 
  }
\end{figure*}

\begin{figure*}[!t]
  \includegraphics[width=\linewidth]{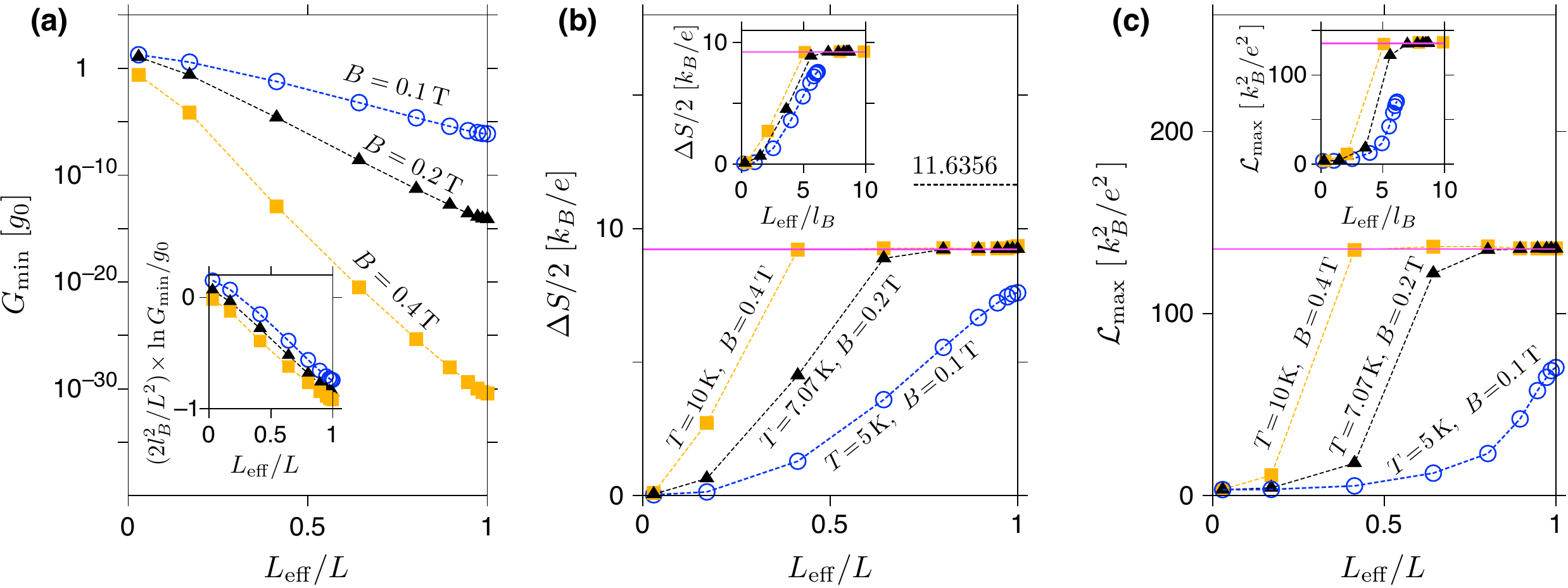}
  \caption{ \label{gmislma3pan:fig}
  (a)--(c) Thermoelectric characteristics of the disk with smooth potential
  barriers in the quantum Hall regime displayed versus $L_{\rm eff}$
  given by Eq.\ (\ref{ldiffm}).
  (a) The minimum zero-temperature conductance reached for $-E_{1{\rm LL}}<E<0$,
  with inset visualizing the scaling according to
  $G_{\rm min}\propto{}\exp(-\frac{1}{2}L_{\rm eff}^2/l_B^2)$ with
  $l_B=\sqrt{\hbar/eB}$ the magnetic length. 
  (b) The maximum thermopower amplitude. 
  (c) The maximum Lorentz number.
  Insets in (b,c) show same data as functions of the rescaled length
  $L_{\rm eff}/l_B$.
  The magnetic field is varied between the datasets (see datapoints, dashed
  lines are guide for the eye only). Additionally, in (b,c) the temperature
  is varied to keep the constant $E_{1{\rm LL}}/(2k_B{}T)\simeq{}11.6356$.
  Solid horizontal lines in (b,c) mark the values following from Eqs.\
  (\ref{smaxsmin}) and (\ref{llnmax}). 
  }
\end{figure*}

The consequences of the above-mentioned features of $G(E)$ for thermoelectric
properties are discussed next.

In Fig.\ \ref{slmpots8pan:fig} we show pairs of plots analogous to those
shown in Fig.\ \ref{smaxlmax2pan:fig}, i.e., the thermopower amplitude
$\Delta{}S=(S_{\rm max}-S_{\rm min})/2$, and the maximum Lorentz number
${\cal L}_{\rm max}$, both displayed as functions of the dimensionless
quantity $E_{1{\rm LL}}/(2k_B{}T)$. 
This time, a~step height is finite and four values of the exponent
$m$ (specified for each panel) are chosen. 
For sufficiently large $E_{1{\rm LL}}/(2k_B{}T)$, the approximations of Eqs.\ 
(\ref{smaxsmin}) and (\ref{llnmax}) [solid lines] are closely followed by
the actual datapoints also for smooth potentials ($m<+\infty$); we further
note that the agreement is generally better for $T=5\,$K [solid symbols]
than for $T=10\,$K [open symbols]. 

The effect of $L_{\rm eff}/L<1$ corresponding to $m<+\infty$,
see Eq.\ (\ref{ldiffm}), and quantifying the
potential smoothness, is further illustrated in Fig.\ \ref{gmislma3pan:fig},
where we show selected thermoelectric properties, $G_{\rm min}$,
$\Delta{}S$, and ${\cal L}_{\rm max}$, as functions of $L_{\rm eff}/L$.
Each dataset corresponds to a~fixed magnetic field ($B=0.1\,$T, $0.2\,$T, or
$0.4\,$T), while the exponent $m$ is varied from $m=2$ to $512$, with
an additional datapoint for rectangular barrier ($m=\infty$)
placed at $L_{\rm eff}/L=1$. 
For finite-temperature characteristics, $\Delta{}S$ and ${\cal L}_{\rm max}$,
we set $T=5\,K$ for $B=0.1\,$T; otherwise,
$T$ is chosen to keep the constant $E_{1{\rm LL}}/(2k_BT)$ ratio (a~quantity 
determining the approximate values $\Delta{}S$ and ${\cal L}_{\rm max}$
via Eqs.\ (\ref{smaxsmin},\ref{llnmax})). 

Remarkably, the datasets for different $B$ reveal some common behavior
upon a~proper rescaling; see three insets in Figs.\ \ref{gmislma3pan:fig}(a),
\ref{gmislma3pan:fig}(b), and \ref{gmislma3pan:fig}(c).
In the first plot, it is easy to see that the conductance away from the
resonances with LLs behaves
approximately as $G_{\rm min}\propto{}\exp(-\frac{1}{2}L_{\rm eff}^2/l_B^2)$.
In the next two plots, the datasets for the finite-temperature characteristics
($\Delta{}S$ and ${\cal L}_{\rm max}$) come much closer to each other if 
plotted as functions of $L_{\rm eff}/l_B$
(where $l_B=\sqrt{\hbar/eB}\simeq{}25.656\,$nm$\,\times{}(B\,[$T$])^{-1/2}$
is the magnetic length) than if simply plotted as functions of $L_{\rm eff}$. 

In addition, the behavior of $G_{\rm min}$ further validates the numerical
stability of the approach presented in Sec.\ \ref{modmet}. Namely, 
the value of $G_{\rm min}/g_0\sim{}10^{-30}$ corresponds to the transmission
amplitude $|t_j|\sim{}10^{-15}$, which coincides with a~typical round-off error
in double precision arithmetic. This is also a~reason why we have
limited our discussion to $B\leqslant{}0.4\,$T (or, equivalently,
$L/l_B\leqslant{}12.33$ for $L=500\,$nm).
For higher $B$, one must use numerical tools employing multiple precision 
\cite{mpack}. In such a~case, a~significant slowdown of the computations
can be expected. 

From a~physical point of view, the considered values of $B$ lead to the
Zeeman splitting of
$\Delta{}E_Z=g\mu_B{}B\approx{}1.16\cdot{}10^{-4}\,$eV$\times{}B\,[$T$]
\ll{}E_{1{\rm LL}}$, with $g\simeq{}2$ and $\mu_B=e\hbar/(2m_e)$
the Bohr magneton.
Throughout this paper the Zeeman term is therefore neglected.

\section{Conclusions}
\label{conclu}

We have investigated selected thermoelectric properties of
graphene-based Corbino disks in the presence of an external magnetic field.
An efficient numerical scheme that allows the determination of these
properties through mode matching for the Dirac equation, up to the magnetic
fields that drive the system into the quantum Hall regime,
using only a~standard double precision arithmetic, is put forward. 

Our results show that both the thermopower amplitude and the maximal
Lorentz number are determined by the energy interval separating the $n=0$ 
and $n=\pm{}1$ Landau levels (LLs) divided by the absolute temperature,
and the fundamental constants. 
(The ratio of the disk radii, as well as the detailed shape of the
electrostatic potential profile, are irrelevant.)
Approximate expressions for the two above-mentioned thermoelectric
characteristics can be derived by assuming the transmission-energy dependence
to be in the form of two Dirac-delta peaks, centered at the energies of
$n=0$ and $n=1$ (or $n=-1$) LLs.
In particular, the expression describing the thermopower amplitude can be
regarded as a~modified version of the well-known Goldsmid-Sharp relation
for semiconductors.
It appears that a~disk-shaped graphene sample, coupled to the two reservoirs
in local thermal equilibrium, can act as a~thermometer measuring the small
temperature difference between the reservoirs (provided that applied
field and average temperature are known). 

Our analysis is carried out within the Landauer-B\"{u}ttiker formalism
for noninteracting quasiparticles. This implies that the fractional quantum
Hall effect (FQHE) is outside the scope of this work.
Although transmission resonances with FQHE states have been observed
in ultraclean
graphene samples \cite{Zen19}, existing thermoelectric measurements for
GaAs/AlGaAs disks \cite{Kob13,Rea20} indicate the presence of integer QHE
states only.
For this reason, a~theoretical study of the thermoelectric signatures
of integer QHE states had to be completed as a~first step.
Undoubtedly, generalizing the approach to include FQHE states would be
a~promising direction for future studies.

\section*{Acknowledgments}
Main part of the work was supported by the National Science Centre of Poland
(NCN) via Grant No.\ 2014/14/E/ST3/00256 (SONATA BIS). 
Computations were performed using the PL-Grid infrastructure.

\section*{Author contributions}
A.R.\ designed the algorithm, A.R.\ and P.W.\ developed the code and performed
preliminary computations, K.R.\ organized the computations on the PL-Grid
supercomputing infrastructure; all authors were involved in data analysis
and manuscript preparation.

\appendix

\begin{figure*}[!t]
  \includegraphics[width=\linewidth]{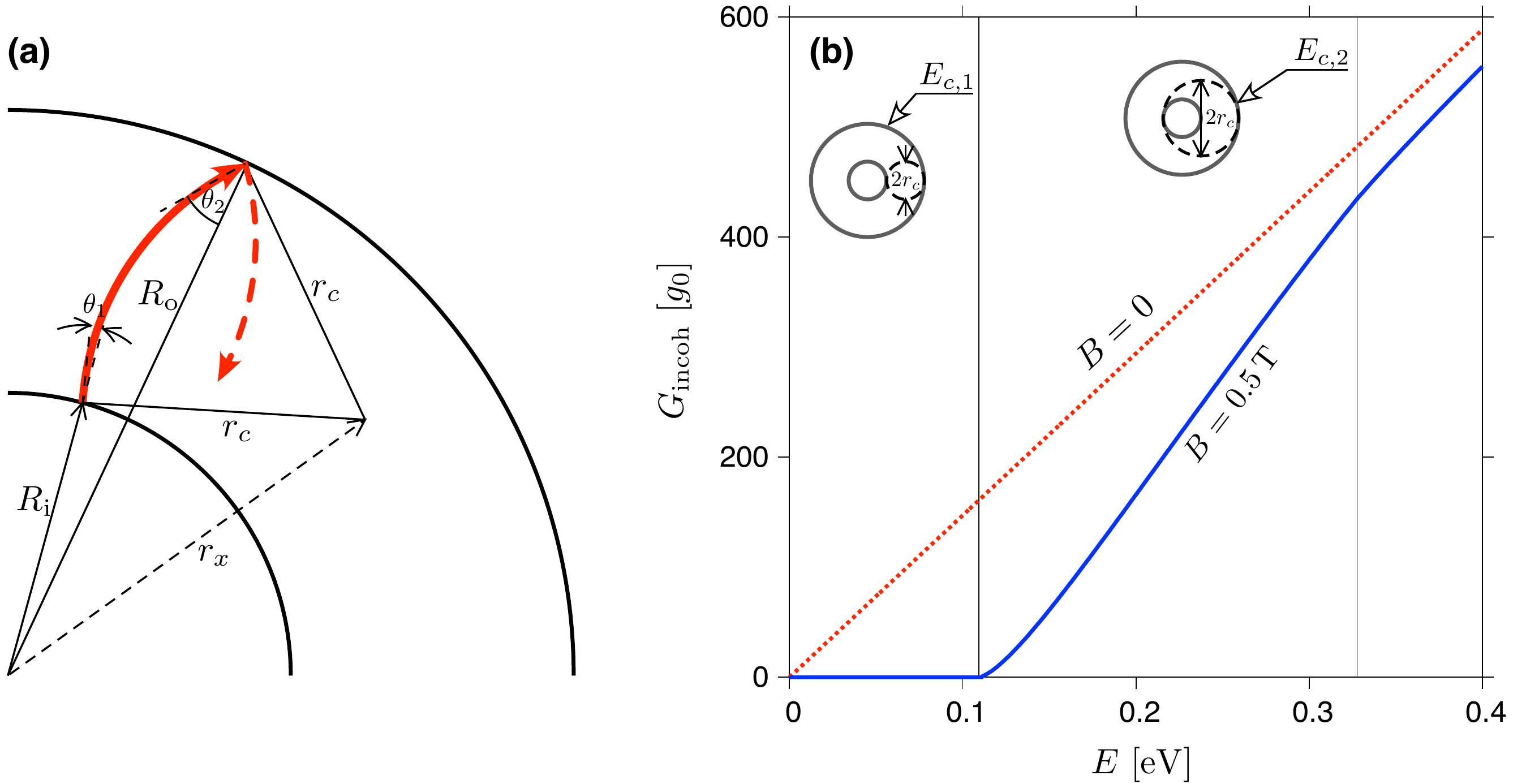}
  \caption{ \label{incohapp:fig}
    (a) Propagation between subsequent scatterings on interfaces at
    $r=R_{\rm i}$ and $r=R_{\rm o}$ (with incident angles $\theta_1$, $\theta_2$)
    in uniform magnetic field, defining the cyclotron orbit centered
    at $r=r_x$ with its radii $r_c$. 
    (b) Incoherent conductance calculated from Eq.\ (\ref{gincoh}) for $B=0$
    [dashed line] and $B=0.5\,$T [solid line]. The disk radii are
    $R_{\rm o}=2R_{\rm i}=1000\,$nm; a~rectangular potential barrier
    is considered. Characteristic Fermi energies $E_{c,1}$ and $E_{c,2}$ 
    mark with vertical lines correspond to $r_c=(R_{\rm o}-R_{\rm i})/2$ and
    $r_c=(R_{\rm i}+R_{\rm o})/2$ (respectively). 
  }
\end{figure*}

\section{Incoherent transport at magnetic field}

Incoherent conductance ($G_{\rm incoh}$) is calculated below by adapting
the method presented in Ref.\ \cite{Ryc22} for the uniform magnetic field
case.

Due to the disk symmetry, the incident angles, $\theta_1$ for the interface
at $r=R_{\rm i}$ and $\theta_2$ for the interface at $r=R_{\rm o}$
(see Fig.\ \ref{incohapp:fig}) remain constant (up to a~sign) during
multiple scattering between the two interfaces. In turn, for a~multimode
regime, summing in Eq.\ (\ref{gtef}) can be approximated by averaging over
the modes, leading to
\begin{equation}
\label{gincoh}
  G_{\rm incoh} = 2g_0k_FR_{\rm i}\left\langle{}T_{12}\right\rangle, 
\end{equation}
where
\begin{equation}
  \label{avtt12}
  \left\langle{}T_{12}\right\rangle = \frac{1}{2} \int_{u_c}^{1}du_1
  \frac{T_1T_2}{T_1+T_2-T_1T_2}, 
\end{equation}
with
\begin{equation}
  T_j = \frac{2\sqrt{1-u_j^2}}{1+\sqrt{1-u_j^2}}, \ \ \ \ j=1,2, 
\end{equation}
being the transmission probabilities for a~potential step of infinite height. 
The parameter $u_1=\sin\theta_1$ is further regarded as an independent
variable.
$u_c$ in Eq.\ (\ref{avtt12}) denotes the minimal value of $u_1$ above which
the corresponding $u_2=\sin\theta_2$ satisfies $|u_2|\leqslant{}1$. 

Assuming a~constant electrostatic potential energy in the disk area, 
the trajectory between subsequent scatterings forms an arc, with the radii
$r_c=|E|/(v_FeB)$ (the cyclotron radius for massless Dirac particle at a~field
$B>0$), centered at the distance $r_x$ from
the origin. Solving the two triangles with a~common edge $r_x$ (dashed line)
and the opposite vertices in two scattering points, one easily finds
\begin{equation}
  r_x^2 = R_{\rm i}^2+r_c^2 + 2R_{\rm i}{}r_c\sin\theta_1 
\end{equation}
(for the triangle containing a~scattering point at $r=R_{\rm i}$), and
\begin{equation}
  r_x^2 = R_{\rm o}^2+r_c^2 - 2R_{\rm o}{}r_c\sin\theta_2 
\end{equation}
(for the triangle containing a~scattering point at $r=R_{\rm o}$),  
leading to  
\begin{equation}
  u_2 = \sin\theta_2 
  = \frac{R_{\rm o}^2-R_{\rm i}^2-2R_{\rm i}r_cu_1}{2R_{\rm o}r_c}. 
\end{equation}
Therefore, the value of $u_c$ in Eq.\ (\ref{avtt12}) is given by
\begin{equation}
  \label{ucases}
  u_c = \begin{cases}
    -1,  & \text{if }\ r_c\geqslant{}\frac{R_{\rm i}+R_{\rm o}}{2} \\
     \frac{R_{\rm o}^2-R_{\rm i}^2}{2R_{\rm i}r_c}-\frac{R_{\rm o}}{R_{\rm i}},
     & \text{if }\
     \frac{R_{\rm i}+R_{\rm o}}{2}>r_c\geqslant{}\frac{R_{\rm o}-R_{\rm i}}{2} \\
    1,  & \text{if }\ r_c<\frac{R_{\rm o}-R_{\rm i}}{2} \\
  \end{cases}. 
\end{equation}

In a~zero-field limit, we have $r_c\rightarrow{}\infty$, leading to
$u_c=-1$ and $u_2\rightarrow{}-(R_{\rm i}/R_{\rm o})u_1$.
In such a~limit, the integral in Eq.\ (\ref{avtt12}) can be calculated
analytically, leading to
\begin{widetext}
\begin{equation}
  G_{\rm incoh}(B\rightarrow{}0) = 2g_0k_FR_{\rm i}
  \frac{(2a+\frac{1}{a})\arcsin{}a+3\sqrt{1-a^2}-\frac{\pi}{2}(a^2+2)}{1-a^2},
  \ \ \ \
  \text{with }\ a=R_{\rm i}/R_{\rm o}. 
\end{equation}
\end{widetext}
The above reproduces a~zero-field result reported in Ref.\ \cite{Ryc22}.
For $B>0$, the integration needs to be performed numerically.

From geometric point of view, the limiting values of $r_c$ in
Eq.\ (\ref{ucases}), i.e., $r_{c,1}=(R_{\rm o}-R_{\rm i})/2$ and
$r_{c,2}=(R_{\rm i}+R_{\rm o})/2$, indicate three distinct situation:
(i) none of the circular trajectories originating from $r=R_{\rm i}$ can reach
$r=R_{\rm o}$ (the $r_c<r_{c,1}$ case),
(ii) trajectories with some incident angles $\theta_1$ may reach the second
interface, but some cannot (the $r_{c,2}>r_c\geqslant{}r_{c,1}$ case), and
(iii) all the trajectories originating from $r=R_{\rm i}$ reach $r=R_{\rm o}$
(the $r_c\geqslant{}r_{c,1}$ case). 

Also in Fig.\ \ref{incohapp:fig}, we display $G_{\rm incoh}$ calculated
numerically from Eq.\ (\ref{gincoh}) for $B=0$ and $B=0.5\,$T, with the
remaining system parameters same as considered throughout the paper.
The Fermi energies $E_{c,\alpha}=v_FeBr_{c,\alpha}$, $\alpha=1,2$, for
the $B=0.5\,$T case, are mark with vertical lines.
It can be shown that for $E\approx{}E_{c,1}$, the incoherent conductance
behaves as
\begin{equation}
  G_{\rm incoh}(E)\propto{}\Theta(E-E_{c,1})\left|E-E_{c,1}\right|^{3/2},  
\end{equation}
with $\Theta(x)$ being the Heaviside step function. 

Remarkably (see the main text), $G_{\rm incoh}$ follows quite close the actual
$G$ calculated via the numerical mode matching, but thermoelectric
characteristics are ruled by the Landau levels, 
with their energies being unrelated to the value of $E_{c,1}$.



\end{document}